\documentclass[12pt]{article}
\usepackage{amsmath,amssymb,graphicx,epic}

\newcommand{\be}{\begin{eqnarray}}
\newcommand{\ee}{\end{eqnarray}}

\def\BX{{\bf x}}
\def\BX0{{\bf x}_0}
\def\a{\alpha}
\def\G{\Gamma}

\def\p{\partial}
\def\d{\delta}
\def\s{\sigma}
\newcommand{\La}{\Lambda}
\newcommand{\Om}{\Omega}
\newcommand{\om}{\omega}

\def\nn{\nonumber}
\def\eps{\epsilon}

\def\E{E_{10}}
\def\K{K(E_{10})}
\def\KE{E_{10}/K(E_{10})}

\def\lae{{\mathfrak{e}}}

\def\bi{\bar{\imath}}
\def\bj{\bar{\jmath}\,}
\def\bk{\bar{k}}

\begin{document}
{\flushright AEI-2006-015\\[1cm]}

\begin{center}
{\bf \Large A note on gauge fixing in\\ Supergravity/Kac--Moody
correspondences}\\[7mm]
Christian Hillmann, Axel Kleinschmidt and Hermann Nicolai \\[5mm]
     {\sl  Max-Planck-Institut f\"ur Gravitationsphysik\\
     Albert-Einstein-Institut \\
     M\"uhlenberg 1, D-14476 Potsdam, Germany}
{}\\[7mm]
\begin{minipage}{12cm}\footnotesize
\textbf{Abstract:} We explain how to achieve the traceless gauge for
the spatial part of the spin connection in the framework of the
recently proposed  correspondence between the (appropriately
truncated) bosonic sectors 
of maximal supergravities and the `geodesic' $\s$-model over $\KE$ at low 
levels. After making this gauge choice, the residual symmetries on both sides
of this correspondence match precisely. The gauge choice also allows 
us to give a physical interpretation
to the multiplicity of certain primitive affine null roots of~$\E$.
\end{minipage}
\end{center}

\vspace{3mm}

Recent work has established intriguing evidence for the realization
of indefinite (sometimes hyperbolic) Kac--Moody algebras in
supergravity and M-theory. In particular, for maximal $D=11$
supergravity \cite{CJS}, there are now several proposals on how to
realize these symmetries. The approach of \cite{We01} seeks a
covariant implementation of the  `very-extended' Kac--Moody algebra
$E_{11}$ via a non-linear realization  directly in eleven dimensions
(possibly augmented by further central  charge coordinates \cite{We03}). 
By contrast, the approach of \cite{DaHeNi02,DaNi04}, based on the 
hyperbolic Kac--Moody algebra $E_{10}$, has its roots in the classic BKL
analysis of  Einstein's equations in the vicinity of a spacelike
(cosmological) singularity~\cite{BKL}, according to which the theory
near the singularity is effectively described by a one-dimensional
reduction, in which spatial  gradients are neglected in comparison
with time derivatives (for a recent review with many references, see
\cite{DaHeNi03}). A `hybrid' approach, combining some of the features
of \cite{We01,DaHeNi02} has been developed in
\cite{EnHo04a,EnHo04b,EnHeHo05}. 

In spite of important conceptual differences between these approaches,
a common feature is that they all require the tracelessness of the 
anholonomicity coefficients (or, equivalently, the spin connection) 
in order to match the (appropriately truncated) degrees of freedom 
between supergravity and the Kac--Moody $\s$-model. For $E_{11}$, 
the issue has been discussed in \cite{We03a}. In this note, we explain
how to realize this gauge in the $\E$-based approach of \cite{DaHeNi02}, 
by making joint use of diffeomorphisms and local Lorentz transformations 
in such a way that, at the end of the gauge fixing procedure, the residual 
symmetries on both sides of the correspondence match precisely. Our 
arguments underline a point already made in \cite{KlNi05a} concerning the
importance of gauge fixing {\em before} making the identification 
between the supergravity theory and the `geodesic' Kac--Moody $\s$-model, 
both at the kinematical and the dynamical level. The traceless gauge 
choice also resolves a puzzle concerning the multiplicity of the 
affine null root (=8 for $\E$) and its images under permutations of
the spatial coordinates; namely, we will show that this multiplicity 
indeed coincides with the number of physically relevant degrees of 
freedom for each choice of null root. In the final section, we comment on 
related issues in the context of the $E_{11}$ proposal of \cite{We01},
and on the extension of the present results to the fermionic sector.

Let us first summarize the basic conjecture and results of \cite{DaHeNi02}. 
As shown in \cite{DaHeNi03}, the relevant equations of motion simplify
near a space-like singularity in the sense that the degrees of freedom
can be divided into `active' ones (the diagonal metric components),
and `passive' ones (off-diagonal metric and various matter degrees
of freedom) which freeze near the singularity. The resulting dynamics 
is thus described by a one-dimensional reduction of the higher dimensional 
field equations (i.e. purely time-dependent equations at a fixed, but
arbitrary spatial point ${\bf x}_0$) which receives effective corrections 
from the passive degrees of freedom (in lowest order in the form of `walls' 
leading to a cosmological billiards).\footnote{It has already been noted 
before that this mechanism offers new possibilities for `emergent spacetime'
scenarios, as the dependence on the spatial degrees of freedom here is 
thought to `emerge out of' (or `disappear into') the spacelike singularity.} 
In the context of supergravity, the possible relevance of a reduction to 
one dimension, and the possible appearance of $\E$ in this reduction, 
had already been foreseen  in \cite{Ju85}, but one crucial difference 
here is that the dependence on the spatial 
coordinates is conjectured to re-emerge via a gradient expansion, which 
gets linked to a level expansion (or height expansion) on the $\s$-model 
side. More precisely, the correspondence is made between the purely 
$t$-dependent $\s$-model degrees of freedom of the Kac--Moody $\s$-model,
and the time-dependent supergravity fields and their (so far only
first order) spatial gradients  
{\em at a fixed spatial point} ${\bf x}_0$.\\

We now explain the successive gauge choices required for the
correspondence of \cite{DaHeNi02}, stressing the residual
symmetries at every step.

{\bf Pseudo-Gaussian gauge}: The analysis of \cite{DaHeNi02} proceeds
from a space-time metric  
in the zero shift (or pseudo-Gaussian) gauge\footnote{For clarity, 
we will stick mostly to $D=11$ supergravity, but the argument remains the same 
for other models of interest in various space-time dimensions $D\leq 11$.} 
\be\label{Gauss}
ds^2 = - N^2 dt^2 + g_{mn} dx^m dx^n \quad , \qquad
N(t,{\bf x}) = n(t) \sqrt{g(t,{\bf x})}
\ee
where indices $m,n,\dots= 1,\dots,10$ label the spatial coordinates,
and $g$ denotes the determinant of the spatial metric, and where the 
purely time-dependent lapse $n(t)$ is to be identified with the one
of the geodesic Kac--Moody $\s$-model, and hence left free. The above 
gauge is supposed to be valid in a tubular neighborhood of the 
worldline parametrized by $\{(t,{\bf x}_0)\,|\, t>0\}$ (in comoving
coordinates). After making this choice, the metric (\ref{Gauss}) is left
invariant by {\em separate}
reparametrizations of the time- and space coordinates, respectively,
that is, $t\rightarrow t'(t)$ and ${\bf x} \rightarrow {\bf x}'({\bf
  x})$, but coordinate changes mixing space- and time coordinates are
disallowed. The pure space reparametrizations are assumed to leave
the point ${\bf x}_0$  invariant (and hence the worldline).

{\bf Vielbein gauge}: Next we make partial use of the local Lorentz
group to bring the elfbein which gives rise to (\ref{Gauss}) into
block-diagonal form. With a (1+10) split of the indices we demand the
form: 
\be\label{10bein}
E_M{}^A = \left(\begin{array}{c|c}N&0\\\hline
0&e_m{}^a\end{array}\right).
\ee
The local space-time Lorentz group $SO(1,10)$ is thereby broken to its
rotation subgroup 
$SO(10)$; that is, (\ref{10bein}) still admits {\em space-time dependent} 
spatial rotations $\La_{ab}(t,{\bf x})$ as a residual symmetry. 

{\bf Traceless spin connection gauge}:
We now wish to exploit this remaining rotation symmetry to set
\be\label{Oabb}
\Om_{ab\,b} (t,{\bf x})=0 \quad\Leftrightarrow\quad
\om_{b\,ba} (t,{\bf x})=0
\ee
where
\be
\Om_{ab\,c} &:=& e_a{}^m e_b{}^n (\p_m e_{nc} - \p_n e_{mc}) = -\Om_{ba\,c}
\nn\\
\om_{a\, bc} &:=& \frac12 \big( \Om_{ab\,c} + \Om_{ca\, b} - \Om_{bc\,a}\big)
     = - \om_{a\, cb} 
\ee
are the spatial components of the coefficients of anholonomicity, and 
the spin connection, respectively. Relation (\ref{Oabb}) is supposed 
to hold in the same tubular neighborhood as (\ref{Gauss}), and implies
the vanishing of the trace and all its spatial gradients along the
the world line $(t,{\bf x}_0)$. The necessity of the tracelessness condition 
arises from the appearance of a representation for the magnetic dual of the 
graviton \cite{Curtright:1980yk,OP,Hull:2001iu,We01,Bekaert:2002uh,We03a} 
at level $\ell=3$ in a level decomposition of $\E$ under its $A_9=SL(10)$
subgroup \cite{DaHeNi02}. The associated tensor of mixed symmetry is
related via the correspondence of ref.~\cite{DaHeNi02} to this dual
graviton by 
\be
P_{a_0|a_1\dots a_8} = \frac32 N \eps_{a_1\dots a_8 bc} \Om_{bc\, a_0}.
\ee
However, from the level decomposition it follows that this 
representation is subject to the irreducibility constraint
\be\label{trcond}
P_{[a_0|a_1\dots a_8]}=0 \quad\Longleftrightarrow\quad
\Om_{ab\,b} = \om_{b\,ba} =0,
\ee
which, as indicated, is equivalent under the dictionary to the
traceless gauge (\ref{Oabb}). Inspection of the available tables of 
higher level representations \cite{NiFi03} reveals the absence of such 
a trace representation at low levels; the relevant representation 
$(000000001)$ appears only at level $\ell=13$, with outer multiplicity 
equal to 22. Similar comments apply to representations corresponding
to the spatial gradients of the trace.

Because both $\Om_{ab\,c}$ and $\om_{a\,bc}$ transform as scalars under
coordinate transformations, it is clear that diffeomorphisms are of no 
further use at this point; in particular, a spatially constant $\Om_{ab\,c}$ 
(with or without trace, e.g. Bianchi cosmologies) remains invariant
under relabeling of the coordinates. 
This is analogous to the traceless gauge $\Gamma^n_{\;nm}=0$ for the 
Christoffel symbol, which transforms as a scalar under local Lorentz 
transformations, whence the role of diffeomorphisms and the local 
Lorentz group is interchanged. Therefore, given a spatial spin connection 
$\om_{a\,bc}$, the problem reduces to solving the equation
\be\label{omtrm}
\om'_{b\,ba} =   \p_b U_{ab} + U_{ab} \om_{c\,cb} = 0
\ee
in terms of the spatial rotation matrix $U_{ab}(t,{\bf x})\in SO(10)$. 
In infinitesimal form (with $V_a\equiv  \om_{b\,ba}$ small, and 
$\p_b U_{ab}=\p_b\La_{ab}$), this equation becomes
\be 
\p_b \La_{ba} = V_a.
\ee
Making the ansatz $\La_{ab}= \p_a v_b - \p_b v_a$, and noticing that
$v_a$ can be chosen divergence-free by shifting $v_a\rightarrow
v_a + \p_a v$ with a suitable $v=v(t,{\bf x})$, we arrive at 
a continuous set of Poisson equations (one for each $t$)
\be\label{poiss}
\triangle v_a (t,{\bf x}) = V_a (t,{\bf x})\;\; ;
\ee
where $\triangle\equiv \p_a\p_a$ is the 10-dimensional spatial
Laplacian. The set of equations (\ref{poiss}) are to be solved in some
tubular neighborhood of the worldline  $(t,{\bf x}_0)$ with
appropriate boundary conditions. The known local existence of
solutions to the Poisson equation guarantees that the gauge
(\ref{Oabb}) can be chosen; moreover the required $SO(10)$ rotation
only fixes the {\em space-dependent} part of the $SO(10)$
transformations since it follows from (\ref{omtrm}) that $\om_{b\,ba}=0$
is not changed by purely {\em time-dependent} $SO(10)$ rotations. \\

{\bf Summary of residual symmetries}:
Having achieved the gauge choices (\ref{Gauss}), (\ref{10bein}) and
(\ref{Oabb}) we are  
left with the following three residual symmetries on the supergravity
side, which can now be directly identified with the residual symmetries
of the $\KE$ $\s$-model in the level decomposition under $A_9$:
\begin{enumerate}
{\item[$(i)$] Reparametrizations of the time parameter $t\rightarrow t'(t)$,
where the time-dependent lapse $n(t)$ in (\ref{Gauss}) is identified 
with the lapse function of the $\KE$ $\s$-model.}
{\item[($ii$)] Purely space-dependent
coordinate transformations (leaving ${\bf x_0}$ inert) that can be
expanded around ${\bf x}_0$ according to
\be
\xi^m ({\bf x}) = \xi^m{}_n ({x}^n - {x}_0^n) + \dots
\ee
The first order term $\xi^m{}_n$ realizes the $GL(10)$ subgroup of the
(global) $\E$. The higher order terms in this expansion are related
to higher order spatial gradients of the various fields, which are 
expected to correspond to higher level representations in the 
decomposition of $\E$ under its $A_9$ subalgebra.\footnote{The relevant
  $\E$ transformations in the $\s$-model will be accompanied by local
  (in time)
  compensating $\K$ transformations. This is analogous to fixing a
  triangular gauge of the spatial vielbein $e_m{}^a$ in (\ref{10bein}).}}
{\item[$(iii)$] Eq.~(\ref{Oabb}) is left invariant
by purely time-dependent spatial rotations $\La_{ab} = \La_{ab}(t)$.
The resulting group $SO(10)$ can be identified with the subgroup of 
$t$-dependent $SO(10)$ rotations within the local `R symmetry'
group $\K$ on the $\s$-model side, which is the finite dimensional
residual invariance left by fixing the triangular gauge for all
fields except in the level $\ell =0$ sector.}
\end{enumerate}
In summary, {\em we have a precise matching not only of the degrees of 
freedom and equations of motion up to level $\ell=3$, but also of the 
residual symmetries on both sides of the correspondence}.\\

Analogous results hold for the $D_9$ and $A_8\times A_1$ decompositions 
\cite{KlNi04a,KlNi04b} of $\E$: one similarly finds no trace representations 
at low levels. For the $A_8\times A_1$ decomposition (corresponding to
IIB, see \cite{KlNi04b}) this is straightforward since one deals
with the dual of the graviton over $A_8=SL(9)$ instead of $SL(10)$, and
the irreducibility constraint (\ref{trcond}) still implies that one
has to fix the space-dependent rotations to arrive at the traceless
gauge. 

{}For $D_9=SO(9,9)$ (related to massive IIA supergravity in
\cite{KlNi04a}) the situation is slightly more involved since the
relevant tensor containing the dual of the graviton is now contained
in an antisymmetric three-form representation of $SO(9,9)$ (at $D_9$
level $\ell=2$), which we denote by $P_{IJK}$ (with $I,J,K =
1,\ldots,18$). Seen from the compact  subgroup $SO(9)\times
\overline{SO(9)}\subset SO(9,9)$ there are four different components
that need to be distinguished ($i,j=1,\ldots,9\, ;\,
\bi,\bj=10,\ldots,18$, cf. \cite{KlNi04a})
\be
\begin{array}{cccccccc}
&P_{ijk} && P_{\bi j k}&& P_{\bi\bj k}&& P_{\bi\bj\bk}\\
SO(9)_{\rm diag}&
\scalebox{.9}{\begin{picture}(30,30)(-10,10)
\thicklines
\multiput(0,0)(10,0){2}{\line(0,1){30}}
\multiput(0,0)(0,10){4}{\line(1,0){10}}
\end{picture}}&&
\scalebox{.9}{\begin{picture}(20,30)(0,5)
\thicklines
\multiput(0,0)(10,0){2}{\line(0,1){20}}
\multiput(0,0)(0,10){3}{\line(1,0){10}}
\put(20,10){\line(0,1){10}}
\multiput(10,10)(0,10){2}{\line(1,0){10}}
\end{picture}}
\oplus
\scalebox{.9}{\begin{picture}(10,30)(10,10)
\thicklines
\multiput(10,10)(10,0){2}{\line(0,1){10}}
\multiput(10,10)(0,10){2}{\line(1,0){10}}
\end{picture}}
&&
\scalebox{.9}{\begin{picture}(20,30)(0,5)
\thicklines
\multiput(0,0)(10,0){2}{\line(0,1){20}}
\multiput(0,0)(0,10){3}{\line(1,0){10}}
\put(20,10){\line(0,1){10}}
\multiput(10,10)(0,10){2}{\line(1,0){10}}
\end{picture}}
\oplus
\scalebox{.9}{\begin{picture}(10,30)(10,10)
\thicklines
\multiput(10,10)(10,0){2}{\line(0,1){10}}
\multiput(10,10)(0,10){2}{\line(1,0){10}}
\end{picture}}
&&
\scalebox{.9}{\begin{picture}(30,30)(-10,10)
\thicklines
\multiput(0,0)(10,0){2}{\line(0,1){30}}
\multiput(0,0)(0,10){4}{\line(1,0){10}}
\end{picture}}\\&
\end{array}
\ee
We have indicated the structure of these four tensors under the
diagonal rotation group $SO(9)_{\rm diag}\subset SO(9)\times
\overline{SO(9)}$. We see that 
those tensors which allow for the mixed symmetry which is required for
(part of) the dual graviton also allow for the presence of a vector
representation. The nine-dimensional trace $\sum_{b=1}^9\om_{b\,ba}$
transforms in a vector representation of $SO(9)_{\rm diag}$ and therefore 
it would seem unnecessary to choose a gauge for it. However, this reasoning
overlooks the dual field for the type IIA dilaton gradient $\p_a\phi$ which 
also transforms as a vector\footnote{$\phi$ is the scalar field
  defined in \cite{KlNi04a} and not strictly identical to the standard
  IIA dilaton.}. Now the appropriate gauge condition relates
the two vectors
\be
\frac12\p_a\phi + \sum_{b=1}^9 \om_{b\,ba} = 0
\ee
Interestingly, this is precisely what the original gauge condition
summed over ten space directions
\be
0=\sum_{b=1}^{10}\om_{b\,ba}= \om_{10\,10a} + \sum_{b=1}^9 \om_{b\,ba}
\ee
translates into if one follows through the redefinitions of
\cite{KlNi04a}. This will be discussed in more detail in \cite{Hi}.

In both cases we see that the matching between supergravity and the
$\KE$ $\s$-model  is possible only if (\ref{Oabb}) is satisfied and
all gauges are fixed so that the residual symmetries agree.\\

{\bf Interpretation of root multiplicity}:
The significance and proper physical interpretation of the imaginary 
roots of $\E$ and their multiplicities in the present context is far from
understood\footnote{Other ideas on the physical r\^ole of imaginary
  roots can be found in \cite{Ganor}.}
 (recall that, generically, imaginary roots $\a$ are
degenerate with exponentially growing multiplicities $\rm{mult}(\a)>1$). 
The above choice of gauge now allows us to extend the matching (and hence 
the `dictionary') beyond real roots, and to give a physical interpretation 
at least for the fact that lightlike (null) roots are associated with root 
multiplicity $>1$. Namely, the roots associated with latter
fall into two classes \cite{DaHeNi03}. First, there are the gravitational
roots (giving rise to `gravitational walls') associated with those
components $\Om_{bc\, a}$, for which the indices $a,b,c$ are all 
different: these correspond to level-3 roots $\a_{abc}$ defined by 
the wall forms (cf. \cite{DaHeNi03}, section~6.2)
\be
\a_{abc} (\beta) = 2\beta^a + \sum_{e\neq b,c} \beta^e
\ee
and are real: $\a_{abc}^2 =2$. The corresponding components of the dual 
field $P_{a_0|a_1\ldots a_8}$ are the ones where $a_0$ is equal to
one of the indices $a_1,\dots, a_8$.

In addition, \cite{DaHeNi03} identified ten subleading gravitational
walls associated with ten null roots, designated as $\mu_a$ for
$a=1,\dots,10$, cf. eqn.~(6.16) there, and defined by the wall forms
\be
\mu_a (\beta) = \sum_{e\neq a} \beta^e
\ee
These ten null roots (for $a=1,\dots,10$) can all be obtained by 
$\mathfrak{sl}(10)$ Weyl reflections (or, equivalently, by permuting 
the spatial coordinates) from the primitive (i.e. lowest height)
null root at height $30$, which has $\d^2=0$, $\rm{mult}(\d)=8$ and is
identical 
to the null root of the affine subalgebra $\mathfrak{e}_9\subset\lae_{10}$
(in the notation of \cite{DaHeNi03}, we have $\d=\mu_1$). This null root,
and its images under the $\mathfrak{sl}(10)$ Weyl group, are the only
imaginary roots appearing on levels $\ell\le 3$ in the $A_9$ decomposition.
The associated components of the dual field $P_{a_0|a_1\ldots a_8}$
belonging to these null roots are the ones for which the indices
$a_0,\dots, a_8$ are all distinct. Using the correspondence we can 
now give a physical interpretation to the multiplicity $\rm{mult}(\d)$.
Since the indices on $P_{a_0|a_1\ldots a_8}$ are all different, two 
indices on the dual coefficient of anholonomicity $\Om_{ab\,c}$ must be 
equal, i.e. we must consider the components\footnote{We
temporarily suspend the summation convention for this discussion, {\em i.e.}
there is {\em no} summation on $b$ here!} $\Om_{ab\,b}$. As shown in
\cite{DaHeNi03},  
these components are then all associated with the null root $\mu_a$, and 
it would thus appear that we have nine possible values for $b$. However, 
thanks to our gauge choice (\ref{Oabb}), there is now one linear relation 
$\sum_{b} \Om_{ab\,b}=0$, whence the number of independent field 
components associated to each null root $\mu_a$ is only eight --- in 
agreement with the root multiplicity $\rm{mult}(\d)=8$! 

How are these statements mirrored in $E_{11}$
\cite{We01,SchnWe01,SchnWe02,We03a}? At least locally, 
the traceless gauge $\Om_{AB}{}^B=0$ (contractions now to be taken with 
the Minkowski metric in eleven dimensions) can be reached by exploiting 
the full local Lorentz group $SO(1,10)$ \cite{We03a}. The difference 
is now that, after gauge fixing, the local Lorentz group has 
been `used up' completely, and there remains no symmetry to identify 
with the $SO(1,10)$ subgroup of the local group $K(E_{11})$, while
the traceless gauge is still compatible with full 11-dimensional
diffeomorphism invariance. A second difference is that a counting 
argument analogous to the one given above would suggest that there 
are now {\em nine} independent components in $\Om_{AB}{}^B=0$ for each
$A$ ({\em no} summation on $B$), whereas the multiplicity of the associated 
null root $\d$ remains the same ($=8$) when $\d$ is considered as a root 
of $E_{11}$. As also mentioned in \cite{We03a}, instead of discarding 
the trace (in order to retain full Lorentz invariance), one might look 
for a trace representation at higher levels. Inspection of the tables 
\cite{NiFi03} reveals that the relevant representation $(0000000010)$ 
does appear in the $A_{10}$ decomposition of $E_{11}$, but only at 
level $\ell=14$, and with outer multiplicity~491.

{\bf Supersymmetric generalization}:
Similar considerations apply to the supersymmetric version of the $\E$
$\s$-model \cite{DaKlNi06,dBHP}. The Kac--Moody model allows only a local
supersymmetry with parameter $\eps(t)$ depending only on
time.\footnote{Of course, one envisages an analogue of the gradient
  conjecture where the space dependence of the $\eps(t,{\bf x})$ is
  encoded in some `higher level' components of an infinite-dimensional
  spinor of $\K$. Here, we discuss only the truncation to the
  unfaithful spinor representation studied in \cite{DaKlNi06,dBHP}.} 
Therefore, we should require on the supergravity side, a similar gauge
conditon on the supergravity fermions involving spatial gradients, which
reduces $\eps(t,{\bf x})$ to purely time-dependent supersymmetry
transformations with parameter $\eps(t)$. The precise form of this
condition is presently unknown but will be schematically of the form
$\p^m \Psi_m = 0$ (where $\Psi_m$ denotes the spatial components of the
gravitino). We note also that one can consider a completely gauge-fixed 
version of the model where one chooses the lapse $n(t)=1$ which is 
reflected in the supersymmetric partner constraint $\psi_0 -
\G_0\G^a\psi_a=0$ \cite{DaKlNi06}.

\vspace*{2mm}

\noindent
{\bf Acknowledgements}: 
This work was partly supported by the European Research and 
Training Network `Superstrings' (contract number MRTN-CT-2004-512194).

\end{document}